\documentclass[sigconf]{acmart}

\AtBeginDocument{%
  }

\PassOptionsToPackage{table,xcdraw}{xcolor}
\usepackage{colortbl}
\usepackage{cleveref}
\usepackage{fontawesome5}
\usepackage{listings}
\usepackage{parcolumns}
\usepackage{color} 
 
\usepackage{amssymb}

\setcopyright{acmcopyright}
\copyrightyear{2018}
\acmYear{2018}
\acmDOI{10.1145/1122445.1122456}

\acmConference[SBES'24]{Brazilian Symposium on Software Engineering}{September 30 -- October 04,
  2024}{Curitiba, PR}

\setcopyright{none} 
\renewcommand\footnotetextcopyrightpermission[1]{} 

\definecolor{codegreen}{rgb}{0,0.6,0}
\definecolor{codegray}{rgb}{0.5,0.5,0.5}
\definecolor{codepurple}{rgb}{0.58,0,0.82}
\definecolor{backcolour}{rgb}{0.95,0.95,0.92}

\lstdefinestyle{mystyle}{
    backgroundcolor=\color{backcolour},   
    commentstyle=\color{codegreen},
    keywordstyle=\color{magenta},
    numberstyle=\tiny\color{codegray},
    stringstyle=\color{codepurple},
    basicstyle=\ttfamily\footnotesize,
    breakatwhitespace=false,         
    breaklines=true,                 
    captionpos=b,                    
    keepspaces=true,                 
    numbers=left,                    
    numbersep=5pt,                  
    showspaces=false,                
    showstringspaces=false,
    showtabs=false,                  
    tabsize=2
}
\newcommand{\gptDetect}{21}
\newcommand{\gptDetectPerc}{70\%{}}
\newcommand{\mistralDetect}{15}
\newcommand{\geminiDetect}{17}

\begin{document}

\title{Evaluating Large Language Models in Detecting Test Smells}

\author{Keila Lucas}
 \orcid{0000-0003-0730-5846}
 \affiliation{%
   \normalsize \institution{Federal University of Campina Grande} \country{Brazil}
 }
 \email{keila.santos@copin.ufcg.edu.br}
 
 \author{Rohit Gheyi}
 \orcid{0000-0002-5562-4449}
 \affiliation{%
   \normalsize \institution{Federal University of Campina Grande} \country{Brazil}
 }
 \email{rohit@dsc.ufcg.edu.br}

 \author{Elvys Soares}
\orcid{0000-0001-7593-0147}
\affiliation{
  \department{Federal Institute of Alagoas} 
  \country{Brazil}
}
\email{elvys.soares@ifal.edu.br}

  \author{Márcio Ribeiro}
 \orcid{0000-0002-4293-4261}
 \affiliation{%
   \normalsize \institution{Federal University of Alagoas} \country{Brazil}
 }
 \email{marcio@ic.ufal.br}

  \author{Ivan Machado}
 \orcid{0000-0001-9027-2293}
 \affiliation{%
   \normalsize \institution{Federal University of Bahia} \country{Brazil}
 }
 \email{ivan.machado@ufba.br}

\begin{abstract}

Test smells are coding issues that typically arise from inadequate practices, a lack of knowledge about effective testing, or deadline pressures to complete projects.
%
The presence of test smells can negatively impact the maintainability and reliability of software. While there are tools that use advanced static analysis or machine learning techniques to detect test smells, these tools often require effort to be used.
%
This study aims to evaluate the capability of Large Language Models (LLMs) in automatically detecting test smells.
%
We evaluated ChatGPT-4, Mistral Large, and Gemini Advanced using 30 types of test smells across codebases in seven different programming languages collected from the literature.
%
ChatGPT-4 identified \gptDetect{} types of test smells. Gemini Advanced identified \geminiDetect{} types, while Mistral Large detected \mistralDetect{} types of test smells.
%
The LLMs demonstrated potential as a valuable tool
in identifying test smells.
\end{abstract}


\maketitle

\section{Introduction}
\label{sec:introduction}

The testing phase in software development is an important process to ensure the quality, functionality, and security of systems~\cite{spadini2018}. 
Researchers have coined the term test smells to describe potential design problems in test code, analogous to the code smells found in poorly designed source code~\cite{van2001refactoring,beck2003}. 
These symptoms can cause tests to exhibit erratic behavior, such as flakiness, false positives, and false negatives, compromising software quality due to their limited defect-catching capabilities. Despite the concept of test smells not being new, some studies have shown that they are prevalent in both open-source and industry projects, and they negatively impact code maintenance and understanding activities~\cite{bavota2015test,van2007detection,Peruma2019Distribution}. \citet{van2001refactoring} and \citet{mesaros} defined catalogs of test smells along with their refactoring actions.

There are various types of test smells, which can be characterized by code duplication (\textit{Duplicated Assert})~\cite{Peruma2019Distribution}, complexity in conditional structures, lack of documentation of assertions (\textit{Assertion Roulette}), non-deterministic execution behavior~\cite{palomba2020}, among others. Test smells can condition future difficulties in the maintenance process~\cite{SoaresAORGSMSFB23}, as the incidence of test smells in the code base will impair the comprehensibility of the implemented structures as the software evolves~\cite{junior2020survey,junior2021}.
Listing~\ref{lst:example-magic} presents an example of the \textit{Magic Numbers} test smell that occurs when a test method contains inexplicable and undocumented numeric values as parameters or values for identifiers.
The numbers \texttt{15.5D}, \texttt{15} and \texttt{30} are magic numbers, since there is no indication of their semantics.

\begin{lstlisting}[basicstyle=\footnotesize, language=Java, label=lst:example-magic, caption=Example code with \textit{Magic Numbers Test.}]
@Test
public void testGetLocalTimeAsCalendar() {
  Calendar localTime = calc.getLocalTimeAsCalendar(BigDecimal.valueOf(15.5D), Calendar.getInstance());
  assertEquals(15,localTime.get(Calendar.HOUR_OF_DAY));
  assertEquals(30,localTime.get(Calendar.MINUTE));
}
\end{lstlisting}

Recent studies~\cite{kim2021, panichella2022, SoaresRGAS23, pontillo2024, Santana2024} have investigated the impact of test smells~\cite{campos2021} and how these indicators of poor coding affect the comprehension and quality of software~\cite{spadini2018}. 
Existing tools that detect test smells often rely on advanced static analysis or machine learning techniques, which can be complex to implement and extend with new smells, as well as support multiple languages~\cite{aljedaani2021test}. Moreover, these tools require effort to use. Only a few provide explanations and suggestions for code improvements.

Artificial Intelligence (AI) techniques, particularly Large Language Models (LLMs), offer the potential to improve the test review process~\cite{wang2024}. LLMs have revolutionized natural language processing, demonstrating remarkable performance across various tasks, including question answering, machine translation, and text generation~\cite{wang2024}. These models have also impacted various domains within software engineering~\cite{hou2024}, expanding the possibilities of integrating natural language analysis capabilities to enhance source code analysis, including detecting test smells.
However, to the best of our knowledge, no study so far indicates to what extent LLMs can help detect test smells.

In this paper, we evaluate the capability of LLMs in the automatic detection of test smells. 
We evaluated ChatGPT-4, Mistral Large, and Gemini Advanced. These models were tested on 30 types of test smells across codebases in seven different programming languages, which were collected from the existing literature~\cite{test-smell-catalog}.
ChatGPT-4 detected \gptDetect{} types of test smells. Gemini Advanced identified \geminiDetect{} types, while Mistral Large detected \mistralDetect{} types of test smells.
The results indicate that LLMs demonstrate efficiency in detecting test smells. ChatGPT-4, in particular, can identify \gptDetectPerc{} of the test smell types in the code, suggesting that LLMs could be a valuable tool in enhancing the quality of software by identifying and mitigating test smells.
The LLMs demonstrated potential as valuable tools to be integrated into IDEs for the detection of test smells. Their processing capabilities enable the detection and explanation of various types of test smells and offer suggestions for code improvements.


\vspace{-0.2cm}
\section{Methodology}
\label{sec:methodology}


\subsection{Research Questions} 

Our goal is to evaluate ChatGPT-4, Gemini Advanced 1.5, and Mistral Large concerning detecting test smells from the point of view of researchers in the context of test smells cataloged in the literature.
We address the following research questions:
\begin{description}
    \item[RQ$_{1}$] To what extent can ChatGPT-4 detect test smells?
    \item[RQ$_{2}$] To what extent can Gemini Advanced detect test smells?
    \item[RQ$_{3}$] To what extent can Mistral Large detect test smells? 
\end{description}
We compare the output of each LLM with the test smells presented in the catalog~\cite{test-smell-catalog}.

\subsection{Planning} 

The Open Catalog of test smells~\cite{test-smell-catalog} provides a dataset of 127 formal and informal sources, featuring various types of test smells, identified by name, Also Known as (AKA) when available, conceptual definition, and code examples, as well as citations of references for further literature review. The catalog consists of six categories of test smells: Code Related, Dependencies, Design Related, Issues in Test Steps, Test Execution -- behavior, and Test Semantic -- logic.

We evaluated a set of 30 test smells from the catalog~\cite{test-smell-catalog}. 
Each test smell is illustrated with at least one small example, some of which are extracted from real projects~\cite{SoaresRGAS23,ease-2024,DBLP:conf/icse/HauptmannJEHVB04}.
In eleven cases, there are \texttt{$\langle$code snippets$\rangle$} of unit tests implemented using the JUnit framework. Three cases correspond to complete programs along with the test class. In four cases, the examples consist solely of the test class. Meanwhile, in the remaining twelve cases, the examples are snippets of individual methods or functions.
Some types (i.e., \textit{Assertion Roulette}, \textit{Duplicate Assert}, \textit{Exception Handling}, and \textit{Test Code Duplication}) were considered due to their frequent occurrence in popular open-source projects~\cite{SoaresRGAS23}.

For the queries, we use zero-shot prompting~\cite{zero-shot-prompt,prompt-techniques,prompts}, which refers to a query scenario in which the machine learning model receives a task for which it has not been explicitly trained to perform. 
The model uses the general understanding of the submitted query and the pre-existing knowledge to perform the task without any additional adjustments or specific examples related to the task in question.
We use the following prompt in each LLM with default parameters:
\begin{itemize}
    \item Consider the following \texttt{$\langle$language$\rangle$} test case. Does it have any test smells? \texttt{$\langle$code snippet$\rangle$}
\end{itemize}
We evaluate test smells in the following \texttt{$\langle$language$\rangle$}: C\#{}, Java, JavaScript, Python, Ruby, Smalltalk and TTCN-3.
The consultation with the LLMs occurred in May 2024. 




\section{Results}
\label{sec:results}

The LLMs showed promising results in identifying test smells in source code. 
ChatGPT-4 demonstrated the best performance in detecting test smells, successfully identifying \gptDetect{} out of 30 types, with the misses being the 
\textit{Bad Comment Rate},
\textit{Duplicated Code In Conditional}, \textit{Duplicate Statements}, \textit{Irrelevant Information} and \textit{Overcommented Test} test smells. It partially detects the \textit{Badly Used Fixture}, \textit{Constant Actual Parameter Value}, \textit{Exception Handling}, and \textit{Two For The Price Of One} test smells.
The Gemini detect \geminiDetect{} test smells and three partially detects, 
while Mistral detect \mistralDetect{} test smells and five partially detects.
At least one LLM can detect 25 out of 30 test smells. 
The \textit{Overcommented Test} smell in Smalltalk and the \textit{Duplicated Code In Conditional} in TTCN-3 are not detected by the evaluated LLMs in our study.

Table~\ref{tab:summary} presents the detailed performance of LLMs in identifying the different types of test smell, with highlights for correct identifications (\checkmark), partial correct identifications (\faAdjust), errors ($\times$), language, and lines of code (LOC).
Partial hits are characterized when the LLM returns information related to the definition of the test smell but identifies it as a different type of test smell, which has a similar definition, or when the presented information is too simple, without details for the example presented. 

\begin{table*}[]
\caption{LLMs performance summary.}
\label{tab:summary}
\small 
\begin{tabular}{llcrccc}

\rowcolor[HTML]{030101} 
{\color[HTML]{FFFFFF} \textbf{ID}} & {\color[HTML]{FFFFFF} \textbf{Test Smell}} & {\color[HTML]{FFFFFF} \textbf{Language}}  & {\color[HTML]{FFFFFF} \textbf{LOC}} & {\color[HTML]{FFFFFF} \textbf{ChatGPT-4}} & {\color[HTML]{FFFFFF} \textbf{Gemini Advanced}}  & {\color[HTML]{FFFFFF} \textbf{Mistral Large}} \\
1  & Anonymous Test                       & Java       & 4   & \checkmark          & \checkmark              & \checkmark     \\
2  & Assertion Roulette                   & Python     & 20  & \checkmark          & \faAdjust               & \checkmark            \\
3  & Asynchronous Test                    & Java       & 17  & \checkmark          & \checkmark              & \checkmark          \\
4  & Bad Comment Rate                     & TTCN-3     & 52  & $\times$           & $\times$               & \checkmark            \\
5  & Badly Used Fixture & Java       & 24 & \faAdjust          & \faAdjust               & \checkmark           \\
6  & Constant Actual Parameter Value      & TTCN-3     & 13  & \faAdjust          & $\times$               & \faAdjust     \\
7  & Context Logic In Production Code     & Java       & 7   & \checkmark          & \checkmark              & \checkmark           \\
8  & Duplicate Assert                     & Java       & 24  & \checkmark    & \checkmark   & \checkmark     \\
9  & Duplicated Code In Conditional       & TTCN-3     & 22  & $\times$           & $\times$               & $\times$     \\
10 & Duplicate Statements                 & TTCN-3     & 9   & $\times$          & \checkmark              & \checkmark           \\
11 & Empty Test                           & Java       & 5   & \checkmark          & \checkmark              & \checkmark           \\
12 & Exception Handling                   & Java       & 30  & \faAdjust          & \checkmark               & $\times$            \\
13 & Expected Exceptions And Verification & Java       & 8  & \checkmark          & \checkmark                & $\times$           \\
14 & Fire And Forget                      & Ruby       & 22 & \checkmark          & \checkmark              & \faAdjust           \\
15 & Hard-Coded Test Data                 & Java       & 9  & \checkmark          & $\times$               & \checkmark            \\
16 & Hidden Test Call                     & C\#       & 17  & \checkmark          & \checkmark               & \faAdjust           \\
11 & Irrelevant Information               & Java       & 11  & $\times$          & $\times$               &  \faAdjust          \\
18 & Long Test                            & Ruby       & 41  & \checkmark          & \checkmark              & \checkmark          \\
19 & Magic Number Test                    & Java       & 6   & \checkmark          & $\times$               &  \checkmark    \\
20 & Obscure Test                         & Ruby       & 18  & \checkmark          & $\times$               & $\times$            \\
21 & Overcommented Test                   & Smalltalk  & 21  & $\times$           & $\times$               & $\times$            \\
22 & Overspecified Tests                  & Java       & 30  & \checkmark          & \checkmark              & \checkmark          \\
23 & Plate Spinning                       & JavaScript & 50 & \checkmark          &  \checkmark        &  \faAdjust   \\
24 & Redundant Print                      & Java       & 9   & \checkmark          & \checkmark              & \checkmark           \\
25 & Self Important Test Data             & Ruby       & 67  & \checkmark          & \checkmark                & $\times$           \\
26 & Self-Test                            & Java       & 13  & \checkmark    & $\times$              & $\times$            \\
27 & Sensitive Equality                 & Java       & 12  & \checkmark          & \faAdjust                & $\times$           \\
28 & Test Code Duplication                & Python     & 18  & \checkmark          & \checkmark              & \checkmark           \\
29 & The First And Last Rites             & Java       & 13  & \checkmark          & \checkmark              & \checkmark           \\
30 & Two For The Price Of One             & Java       & 13  & \faAdjust          & $\times$              & $\times$  \\ \hline
\multicolumn{4}{r}{\textbf{Total}}            & \gptDetect           & \geminiDetect                & \mistralDetect \\ \hline

\multicolumn{7}{l}{\checkmark hits \hspace{0.2cm} \faAdjust \hspace{0.05cm} partial hits \hspace{0.2cm} $\times$ errors}
\end{tabular}
\end{table*}


\section{Discussion}
\label{sec:discussion}


\subsection{Number of Attempts}

Due to the probabilistic nature of their processing, the models can produce varied responses to the same query, even when the same prompt is applied~\cite{threats-llms-icse-nier-2024}.
For the test smells not identified in the first submission, we execute two additional attempts (2$^{nd}$ and 3$^{rd}$) for each specific example, using a unique chat session for each round with the same prompt. When the LLM succeeded on the second attempt, a third attempt was not necessary. 
The types of test smells that were detected on the first query attempt were not subjected to additional attempts. All correct responses obtained in the outputs were reviewed by two authors to formally record the performance of each LLM.
In the additional attempts (2$^{nd}$ and 3$^{rd}$), no feedback was provided on the results presented, all additional attempts were made by resetting the conversation window, and applying the same prompt.
Table~\ref{tab:summary-attempts} presents the overall summary of the attempts executed on ChatGPT-4, Gemini Advanced, and Mistral Large. 


In response to our \textbf{RQ$_{1}$}, the results of the analysis confirm that ChatGPT-4 demonstrated the best detection performance over 3 attempts, identifying 26 out of 30 examples in total. ChatGPT-4 was unable to detect the \textit{Duplicated Code in Conditional} and \textit{Overcommented Test} test smells, and only partially detected the \textit{Constant Actual Parameter Value} and \textit{Two for the Price of One} test smells.
The results regarding the performance of Gemini Advanced address our \textbf{RQ$_{2}$}, in which we highlight that the LLM identified 17 types of test smells on the first query attempt. In additional attempts (2$^{nd}$ and 3$^{rd}$), Gemini Advanced showed improvement by identifying seven more types of the 13 that were not correctly identified in the first attempt. Table~\ref{tab:gemini-attempts} presents the attempts for the test smells re-evaluated in Gemini Advanced.

Mistral Large had the weakest performance in the test smell detection process, and in response to our \textbf{RQ$_{3}$}, we observed that the LLM correctly detected only 15 types of test smells on the first query attempt. In the additional attempts (2$^{nd}$ and 3$^{rd}$), the LLM was able to identify six more types of test smells that were previously undetected, specifically: \textit{Expected Exceptions And Verification}, \textit{Hidden Test Call}, \textit{Obscure Test}, \textit{Plate Spinning}, \textit{Self-Test}, and \textit{The First And Last Rites}.

\begin{table}[h!]
\caption{Number of test smells identified per attempt for each LLM.}
\small  
\label{tab:summary-attempts}
\begin{tabular}{|l|c|c|c|c|}
\hline
\rowcolor[HTML]{343434} 
{\color[HTML]{FFFFFF} \textbf{LLM}} & {\color[HTML]{FFFFFF} \textbf{1$^{st}$}} & {\color[HTML]{FFFFFF} \textbf{2$^{nd}$}} & {\color[HTML]{FFFFFF} \textbf{3$^{rd}$}} & {\color[HTML]{FFFFFF} \textbf{Total}} \\ \hline
ChatGPT-4  & 21/30  &  3/9  &  2/6  & 26/30 \\ \hline
Gemini Advanced  & 17/30  &  5/13  &  2/8  & 24/30 \\ \hline
Mistral Large  & 15/30  &  5/15  &  1/10  & 21/30 \\ \hline
\hline
\end{tabular}
\end{table}

All LLMs do not detect the \textit{Duplicated Code In Conditional} test smell (see Listing~\ref{lst:example-duplicate-code}) in 3 attempts. 
ChatGPT detected other types of issues in the functions \texttt{checkSomething} and \texttt{checkSomethingElse}, for example, \textit{Magic Numbers}, \textit{Multiple Exit Points}, \textit{External Dependencies} and \textit{Implicit Else Condition}. 
In each attempt, the ChatGPT output shows little variation in responses. 
\begin{lstlisting}[basicstyle=\footnotesize,  label=lst:example-duplicate-code, language=Java, caption=Example code with \textit{Duplicated Code In Conditional.}]
function checkSomething(in float p1, in float p2) return boolean {
    if (p1 < 0.0) {
        return false;
    } if (p2 >= 7.0) {
        return false;
    } ...
}
function checkSomethingElse(in float p1) runs on ExampleComponent {
    var charstring result;
    if (p1 > 0) {
        result := "foo";
        pt.send(result);
    } else {
        result := "bar";
        pt.send(result);
    }
}
\end{lstlisting}


\begin{table}[h!]
\caption{Gemini attempts.}
\label{tab:gemini-attempts}
\small  
\begin{tabular}{|l|c|c|c|}
\hline
\rowcolor[HTML]{343434} 
{\color[HTML]{FFFFFF} \textbf{Test Smell}} & {\color[HTML]{FFFFFF} \textbf{1$^{st}$}} & {\color[HTML]{FFFFFF} \textbf{2$^{nd}$}} & {\color[HTML]{FFFFFF} \textbf{3$^{rd}$}}  \\ \hline
Assertion Roulette  & \faAdjust  & $\times$ & \checkmark \\ \hline
Bad Comment Rate  & $\times$ & \faAdjust & \checkmark  \\ \hline
Badly Used Fixture  & \faAdjust & \checkmark &  \\ \hline
Constant Actual Parameter Value  & $\times$ & $\times$ & $\times$ \\ \hline
Duplicated Code In Conditional  & $\times$ & $\times$ & $\times$ \\ \hline
Hard-Coded Test Data  & $\times$ & \checkmark & \\ \hline
Irrelevant Information  & $\times$ & $\times$ & $\times$ \\ \hline
Magic Number Test  & $\times$ & \checkmark &  \\ \hline
Obscure Test  & $\times$ & \checkmark &  \\ \hline
Overcommented Test  & $\times$ & $\times$ & $\times$ \\ \hline
Self-Test  & $\times$ & $\times$ & $\times$ \\ \hline
Sensitive Equality & \faAdjust & \checkmark &  \\ \hline
Two For The Price Of One  & $\times$ & $\times$ & $\times$ \\ \hline
\hline
\end{tabular}
\end{table}

\subsection{Detecting Test Smells}

Six test smells are detected by only one LLM in the first attempt: \textit{Bad Comment Rate} (Mistral), \textit{Badly Used Fixture} (Mistral), \textit{Exception Handling} (Gemini), \textit{Obscure Test} (ChatGPT), Self-Test (ChatGPT) and \textit{Sensitive Equality} (ChatGPT), as presented in Table~\ref{tab:summary}.
Four test smells are not detected by any LLM (the \textit{Constant Actual Parameter Value}, \textit{Duplicated Code In Conditional},  \textit{Overcommented Test}, and \textit{Two For The Price Of One} test smells). 

LLMs also suggest other test smells in the code snippets evaluated.
Some responses from the LLMs did not exactly match the label of the test smell recorded in the catalog~\cite{test-smell-catalog} but provided related and coherent explanations. 
For instance, the \textit{Anonymous Test} smell occurs when a test whose name is meaningless as it does not express the purpose of the test in the current context~\cite{test-smell-catalog}. 
Consider the example presented in Listing~\ref{lst:anonymous-test}.
ChatGPT-4 yields the following output considering it as the \texttt{Inadequate Test Method Name} test smell:
\begin{itemize}
    \item ``The name of the test method, \texttt{test1}, is not descriptive. It does not convey what aspect of the \texttt{LoginPage.login} method it is testing. Good test names should describe the purpose of the test and the specific behavior or scenario being tested.''
\end{itemize}
Since the test smells definitions are similar, we consider that the ChatGPT-4 answer is correct. 
In some cases, LLMs also suggest refactorings to remove test smells, such as providing better names.
ChatGPT-4 also indicates other problems in the test case. For example, it indicates that the test case does not have assertions.

\begin{lstlisting}[basicstyle=\footnotesize,  label=lst:anonymous-test, language=Java, caption=Example code with \textit{Anonymous Test}.]
@Test
public void test1() {
  LoginPage.login("user", "password");
} 
\end{lstlisting}


The \textit{Assertion Roulette} test smell occurs when a test method has multiple non-documented assertions. It is not possible to understand the reason for the failure of the test~\cite{test-smell-catalog}. 
Listing~\ref{lst:example-assertion-roulette} presents part of the code snippet containing this test smell.
ChatGPT-4 and Mistral detected this test smell on the first attempt, while Gemini provided a partially correct response, indicating that there is an ``Assertion Overload'' in the code example.
Mistral correctly yields the following explanation:
\begin{itemize}
    \item ``Lack of Assertion Messages: The test case does not provide any custom assertion messages. This can make it harder to understand why a test failed when it does. Providing meaningful assertion messages can make the test results more readable.''
\end{itemize}
In the first two attempts, Gemini does not complain about excessive assertions. However, in the third attempt, it identifies the test smell:
\begin{itemize}
    \item ``Excessive Assertion: The test asserts multiple things at once (\texttt{mileage}, \texttt{fullFuel}, \texttt{isValidAirLineCode}). This violates the one assertion per test principle, making it harder to pinpoint the cause of a failure.''    
\end{itemize}

\begin{lstlisting}[basicstyle=\footnotesize,  label=lst:example-assertion-roulette, language=Python, caption=Example code with \textit{Assertion Roulette.}]
class TestFlight(unittest.TestCase):
  def test_flight(self):
    flight = Flight('2569',1000)
    self.assertEqual(flight.mileage,1000)
    self.assertTrue(flight.fullFuel)
    self.assertTrue(flight.isValidAirLineCode())
\end{lstlisting}

The \textit{Exception Handling} smell occurs when a test method explicitly passes or fails and is dependent on the production method throwing an exception. Developers should utilize JUnit's exception handling to automatically pass/fail the test instead of writing custom exception handling code or throwing an exception.
Listing~\ref{lst:example-exception-handling} presents part of the code snippet containing this test smell.
Gemini correctly detects it in the first attempt:
\begin{itemize}
    \item ``Lack of Error Handling Assertions: The test does not verify the correct handling of error conditions or exceptional cases, which can leave gaps in test coverage.''
\end{itemize}
ChatGPT-4 partially detects the problem in the \texttt{Assert.fail} but it does not provide a good explanation:
\begin{itemize}
    \item ``Assertion Messages: The test uses \texttt{Assert.fail(e.getMessage());} within a try-catch block to handle exceptions, which is a valid approach, but the test assertions themselves do not include custom failure messages. This could make it harder to diagnose issues because the output will not explain what the expected logic or outcome was supposed to be, only that the expected and actual values did not match.''
\end{itemize}
However, it correctly identifies the test smell in the third attempt:
\begin{itemize}
    \item ``Error Handling in Tests: The use of try-catch block with \texttt{Assert.fail()} inside the catch is a valid technique for ensuring exceptions are not expected. However, this could be improved by using \texttt{@Test(expected = CalculationException.class)} if the intention was to assert that an exception should not be thrown. If the test's intention is to handle an unexpected exception, then it is generally set up correctly.''
\end{itemize}
Mistral does not explain any issues related to \texttt{Assert.fail} in three attempts.

\begin{lstlisting}[basicstyle=\footnotesize,  label=lst:example-exception-handling, language=Java, caption=Example code with \textit{Exception Handling.}]
@Test
public void realCase() {
  Point p34 = new Point("34", ...); ...
  try {
    a.compute();
  } catch (CalculationException e) {
    Assert.fail(e.getMessage());
  } ...
}
\end{lstlisting}

\subsection{Programming Languages}

The accuracy rate of the LLMs for detecting different types of test smells across various programming languages was evaluated based on the results presented in 
Table~\ref{tab:summary-llms-languages}. Seventeen code examples submitted to the LLMs are written in Java. ChatGPT-4 stands out by correctly identifying 13 (76\%{}) of the 17 types of test smells analyzed.
For other languages, such as C\#, JavaScript, Python, Ruby, and TTCN-3, the LLMs demonstrated varying degrees of success. 
For C\#, ChatGPT-4 and Gemini Advanced detected the \textit{Hidden Test Call} test smell, respectively. 
In Python, Mistral Large and ChatGPT-4 achieved a 100\%{} detection rate (2/2), while Gemini Advanced correctly identified 1 out of 2 test smells. 
For JavaScript, Mistral Large is unable to identify the type of smell in the example, but ChatGPT and Gemini were successful. 
For Ruby, ChatGPT-4, Gemini Advanced, and Mistral Large identified 4/4, 3/4, and 1/4 test smells, respectively. 
For the TTCN-3 language, ChatGPT-4 failed to identify the \textit{Bad Comment Rate}, \textit{Duplicated Code In Conditional}, \textit{Duplicate Statements} test smells on the first attempt. The model achieved only a partial success for the \textit{Constant Actual Parameter Value} test smell, indicating ``Similar test data'' for the function \texttt{f} that sends messages with almost identical data in two cases, but it was not specific about the use of parameters in the test.

All three models failed to detect the \textit{Overcommented Test} smell in three attempts. No LLM was able to identify the issue of excessive comments in Smalltalk, which obscure the code and divert attention from the purpose of the test. The responses either did not reference problems related to comments or diverged from the definition of the submitted test smell type, indicating a detection error.
But they detected other issues, such as the \textit{Mystery Guest}, \textit{Conditional Complexity}, \textit{Test Dependencies}, and \textit{Eager Test} smells.

\begin{table}[h!]
\caption{Number of test smells identified by programming language for each LLM.}
\label{tab:summary-llms-languages}
\small  
\begin{tabular}{|c|c|c|c|}
\hline
\rowcolor[HTML]{343434} 
{\color[HTML]{FFFFFF} \textbf{Language}} & {\color[HTML]{FFFFFF} \textbf{GPT-4}} & {\color[HTML]{FFFFFF} \textbf{Gemini Advanced}} & {\color[HTML]{FFFFFF} \textbf{Mistral Large}}\\ \hline
C\#     & 1/1       & 1/1         & 0/1\\ \hline
Java     & 13/17       & 10/17         & 10/17\\ \hline
JavaScript     & 1/1       & 1/1         & 0/1\\ \hline
Python     & 2/2       & 1/2         & 2/2\\ \hline
Ruby     & 4/4       & 3/4         & 1/4\\ \hline
Smalltalk    & 0/1       & 0/1        & 0/1\\ \hline
TTCN-3    & 0/4       & 1/4        & 2/4\\ \hline
\hline
\end{tabular}
\end{table}

\subsection{Metamorphic Testing}

Metamorphic testing (MT) is an active area of research focused on improving model robustness~\cite{metamorphic-testing-2}. This testing approach involves generating new data samples (code) by applying metamorphic transformations to the original validation or testing data. The modifications applied to the code snippets are controlled to maintain semantic and behavioral equivalence with the original code, while developing modifications in their Abstract Syntax Trees (ASTs), considering the restructuring of variables, parameters, and methods, as well as the removal of comments. The purpose of this methodology is to test the resilience of models, allowing the identification of hidden faults that may not be apparent in the original evaluation.

We select 10 test smells of different sizes and languages. The selected examples are syntactically modified to be submitted in new queries to the LLMs. 
For example, we changed variable names, method names, some numeric parameters, and declared strings. 
The modification made to the original code (Listing~\ref{lst:example-original-code}) of the \textit{Context Logic In Production Code} test smell is presented in Listing~\ref{lst:example-mt-code}. 

\begin{lstlisting}[basicstyle=\footnotesize,  label=lst:example-original-code, language=Java, caption=Original code.]
public static void SaveToDatabase(Customer customerToWrite) {
  if(AreWeTesting)
    WriteWithMockDatabase(customerToWrite);
  else
    Write(customerToWrite);
}
\end{lstlisting}

\begin{lstlisting}[basicstyle=\footnotesize,  label=lst:example-mt-code, language=Java, caption=Listing~\ref{lst:example-original-code} modified for
the metamorphic test.]
public static void TestBD(Customer customerTW) {
  if(makingTest)
    WriteWithMockDatabase(customerTW);
  else
    Write(customerTW);
}
\end{lstlisting}

Table~\ref{tab:mt-summary} presents the performance of the models.
ChatGPT-4 again achieved the best performance, correctly identifying all examples, even for the \textit{Duplicate Statements} test smell that was detected in the second attempt in the original code.
The MT allowed Mistral to correctly detect the test smell \textit{Plate Spinning}, which had previously been detected only partially, as the LLM did not indicate that the test could fail before the calls were completed. After the submission of the MT code, the model identifies that the test can fail due to the unpredictability of external requests.

\begin{table*}[]
\caption{Metamorphic Testing summary.}
\label{tab:mt-summary}
\small  
\begin{tabular}{lcccccc}
& \multicolumn{2}{c}{\cellcolor[HTML]{000000}\color[HTML]{FFFFFF} \textbf{ChatGPT-4}}                                                                    & \multicolumn{2}{c}{\cellcolor[HTML]{000000}\color[HTML]{FFFFFF} \textbf{Gemini Advanced}}                                                            & \multicolumn{2}{c}{\cellcolor[HTML]{000000}\color[HTML]{FFFFFF} \textbf{Mistral Large}}                                                              \\ \hline
\rowcolor[HTML]{EFEFEF} 
{\color[HTML]{030101} \textbf{Test Smell}}                 & \multicolumn{1}{c}{\cellcolor[HTML]{EFEFEF}Original} & \multicolumn{1}{c}{\cellcolor[HTML]{EFEFEF}MT} & \multicolumn{1}{c}{\cellcolor[HTML]{EFEFEF}Original} & \multicolumn{1}{c}{\cellcolor[HTML]{EFEFEF}MT} & \multicolumn{1}{c}{\cellcolor[HTML]{EFEFEF}Original} & \multicolumn{1}{c}{\cellcolor[HTML]{EFEFEF}MT} \\ \hline
Anonymous Test   & \checkmark      & \checkmark          & \checkmark  & \faAdjust  & \checkmark          & \checkmark              \\
Asynchronous Test     & \checkmark    & \checkmark       & \checkmark    & \checkmark            & \checkmark     & \checkmark       \\
Context Logic In Production Code      & \checkmark     & \checkmark     & \checkmark      & \checkmark     & \checkmark    & \checkmark      \\
Duplicate Statements   & $\times$   & \checkmark    & \checkmark      & $\times$  & \checkmark   & \checkmark                       \\
Fire And Forget   & \checkmark     & \checkmark     & \checkmark      & \faAdjust      & \faAdjust    & \faAdjust                     \\
Hard-Coded Test Data    & \checkmark      & \checkmark     &    $\times$       & $\times$   & \checkmark       & \checkmark            \\
Long Test    & \checkmark    & \checkmark      & \checkmark         & \faAdjust  & \checkmark     & \checkmark        \\
Plate Spinning        & \checkmark     & \checkmark     & \checkmark      & \checkmark  & \faAdjust     & \checkmark                     \\
Test Code Duplication     & \checkmark   & \checkmark     & \checkmark      & \checkmark     & \checkmark        & \checkmark            \\
The First And Last Rites     & \checkmark     & \checkmark     & \checkmark    & \faAdjust   & $\times$   & \faAdjust                    \\ \hline 
\multicolumn{1}{r}{\textbf{Total}} &  9     & 10   & 9    & 4  & 7       & 8                                                       \\ \hline
\end{tabular}
\end{table*}

Gemini correctly detects 4 out of 10 test smells in the MT.
It fails to detect the \textit{Duplicate Statements} test smell, not being able to identify the repeated structures. The LLM highlights other issues such as ``Obscure Intent'', ``Conditional Test Logic'', and ``Potential Assertion Roulette'', but it does not emphasize problems related to the repetition of structures in the original code. 
Gemini occasionally misidentified certain types of test smells. The LLM presented the \textit{General Fixture} test smell as the answer for the \textit{Long Test} and \textit{The First And Last Rites} test smells that were identified in the original code.
The confusion between the types \textit{The First And Last Rites} and \textit{General Fixture} may have been caused by the similarity between the definitions of both types. As both types involve problems with the occurrence of recurring structures, the LLM may have interpreted them as synonymous types of smells. 

\vspace{-0.3cm}
\subsection{Threats to Validity}

There are some threats to validity that could impact the results and interpretations~\cite{threats-llms-icse-nier-2024}.
The test smells may be part of the LLMs training data. 
We conducted metamorphic testing to reduce this threat to validity.
The way prompts are structured may influence the responses, potentially making them more general for types that require specific explanations. This could affect the accuracy and specificity of the test smell detection.
We use a simple prompt.
It is not always an easy task to check whether the LLM output is aligned with the test smell definition. We checked each answer with two authors of the paper.

Most test smells are evaluated using a single test case example. 
The limited number of examples per test smell could affect the robustness of the findings. Additionally, some examples are separated from the original class context, which might not fully represent real-world scenarios.
The study focused on a specific set of test smells and evaluated examples primarily in Java, with fewer examples from other languages. This limited diversity may affect the generalizability of the findings to other programming languages and test smell types. Moreover, the study's findings might not apply to other codebases or test cases not represented in the catalog.
The reproducibility of the study results is dependent on the consistent behavior of LLMs, which may vary with updates or changes in the models.


\vspace{-0.4cm}
\section{Related Work}
\label{sec:related}

\citet{pontillo2024} proposed a method based on machine learning (ML) to detect test smells, focusing on four specific test smells: \textit{Eager Test, Mystery Guest, Resource Optimism}, and  \textit{Test Redundancy}, aiming to overcome the limitations of existing heuristic techniques. The authors applied the ML approach to predict the likelihood of a test case being affected by a specific smell. They will also compare their method with established heuristic techniques.

\citet{soares2020Refactoring} aimed to investigate developers' awareness of test smells' existence and acceptance of their refactorings in submitted pull requests. The authors demonstrate that developers are not always acquainted with the terminology of test smells but recognize their effects and harmfulness when consulted. \citet{SoaresRGAS23} conducted a mixed-methods analysis involving 485 Java projects, investigating the extent to which developers adopt JUnit 5 and its new features to enhance test code quality. The study identified JUnit 5 features that can help remove test smells, such as \textit{Assertion Roulette}, \textit{Test Code Duplication}, and \textit{Conditional Test Logic}. 

\citet{aljedaani2021test} compiled a catalog of test smell detection tools. \citet{lambiase2018just} proposed an IntelliJ plugin called DARTS (Detection and Refactoring of Test Smells) that detects the \textit{Eager Test}, \textit{General Fixture}, and \textit{Lack of Cohesion of Test Methods} test smells. Another tool is the RTj framework by \citet{MartinezRTj2020} that deals with \textit{Rotten Green Test Cases}, which are tests that pass despite having at least one unexecuted assertion. Although RTj's test smells are outside our scope, it offers refactoring actions limited to substituting a failing assertion with a call to the \texttt{fail} method and adding a \texttt{TODO} comment to the problematic code. \citet{SantanaRAIDE2020} proposed RAIDE, which is an open-source, IDE-integrated tool that addresses the \textit{Assertion Roulette} and \textit{Duplicated Assert} test smells in Java projects.


\citet{DeBleser2019} assessed the diffusion and perception of test smells in SCALA projects, finding low diffusion of test smells across SCALA test classes and that many developers struggle to correctly identify most smells, despite recognizing design issues. \citet{Peruma2019Distribution} investigated test smells in open-source Android applications and found that developers generally recognize the proposed smells as bad programming practices in unit test files.

\citet{junior2020survey} investigated the causes of test smell introduction by developers, revealing that even experienced professionals introduce test smells during their daily programming tasks despite using standardized company practices. Another study by \citet{spadini2020} examined the severity rating of four test smells and their perceived impact on test suite maintainability. They found that developers consider current detection rules for specific test smells too strict and that the newly defined severity thresholds align with participants' perceptions of how test smells impact test suite maintainability.
\citet{10.1145/3631973} identified and defined new test smell types from software practitioners' discussions on Stack Overflow and developed a detector to identify these smells in real-world Java projects. Through empirical evaluation and practical validation, the study demonstrated the prevalence and impact of these test smells, providing insights for improving test code quality.

In our work, we investigate the extent to which LLMs can be useful for detecting test smells in the source code of software test cases. ChatGPT-4 successfully detects \gptDetectPerc{} of the test smells, showing promising results. In future work, we intend to evaluate to what extent LLMs can help in refactoring test smells.
\vspace{-0.3cm}
\section{Conclusion}
\label{sec:conclusion}

In this paper, we evaluated the potential of LLMs for detecting test smells in source code. The LLMs demonstrated competence in detecting various types of test smells, proving to be an important auxiliary resource for software testing tasks. Notably, ChatGPT-4 achieved the highest overall accuracy rate of \gptDetectPerc{}. All data from this work are available online~\cite{artefatos}.
The results indicate opportunities to enhance tools for detecting test smells. 
By integrating LLMs into the software development lifecycle, developers can more effectively identify and address test smells. The automation of test smell detection using LLMs reduces the manual effort required in the testing phase. 
Further research and development are needed to improve the robustness and accuracy of LLMs in detecting a wider range of test smells, particularly in less common programming languages and more nuanced test smell types.

In future work, we intend to deepen the analysis of test smell types by considering a larger number of examples in test cases across various programming languages. This approach will allow for a more robust evaluation of the performance of ChatGPT-4, Mistral, and Gemini. Additionally, we plan to expand the review to include code bases from open-source projects and to investigate the potential of other LLMs, such as ChatGPT4-o, Claude, and Llama. Gemini has a 1 million-token context window, which enables it to evaluate larger programs. 


\bibliographystyle{ACM-Reference-Format}
\bibliography{referencias}

\end{document}